\documentclass{article}
\usepackage{nips2011,times}
\usepackage{graphics}

\title{Personalized Academic Research Paper Recommendation System\thanks{We sincerely appreciate to Guy Lebanon, Gregory Abowd, and Ling Liu for participating our user study.}}

\author{
Joonseok Lee, Kisung Lee, Jennifer G. Kim \\
Georgia Institute of Technology\\
Atlanta, GA 30318 \\
\texttt{\{jlee716,kslee,jkim693\}@gatech.edu} \\
 }

\begin{document}

\maketitle

\begin{abstract}
A huge number of academic papers are coming out from a lot of
conferences and journals these days. In these circumstances, most
researchers rely on key-based search or browsing through proceedings
of top conferences and journals to find their related work. To ease
this difficulty, we propose a Personalized Academic Research Paper
Recommendation System, which recommends related articles, for each
researcher, that may be interesting to her/him. In this paper, we
first introduce our web crawler to retrieve research papers from the
web. Then, we define similarity between two research papers based on
the text similarity between them. Finally, we propose our
recommender system developed using collaborative filtering methods.
Our evaluation results demonstrate that our system recommends good
quality research papers.
\end{abstract}

\section{Introduction}

Recommender systems are widely used these days in e-commerce, for
the purpose of personalized recommendation. Based on each user's
profile, previous purchase history, and online behavior, they
suggest products which they are likely to prefer. For example,
Amazon.com is using recommender systems for books. When a user
logs-in to the system, it suggests books similar to previously
bought ones by the user.

Personalized recommendation can be applied to outside of commercial
applications. These days, many academic papers are coming out from a
lot of conferences and journals. Academic researchers should go
through all the conferences and journals which are related to their
field of research and find out if there is any new articles that may
relate to their current works. Sometimes they search the articles
from Google scholars or Citeseer with the key words that might show
interesting articles to them. However, these two methods require
users to commit their time to search articles, which is
labor-intensive, and also do not guarantee that they will find the
exact articles related to their field of research.

In order to reduce their workload, we suggest developing the
scholarly paper recommendation system for academic researchers,
which will automatically detect their research topics they are
interested in and recommend the related articles they may be
interested in based on similarity of the works. We believe this
system will save the researchers' time to search the articles and
increase the accuracy of finding the articles they are interested
in.

\section{Related Work}
In this section we briefly present some of the research literature
related to recommender systems in general, academic paper
recommendation system, and evaluation of recommender systems.

Recommender systems are broadly classified into three
categories\cite{jannach:rs}: collaborative filtering, contents-based
methods, and hybrid methods. First, collaborative filtering uses
only user-item rating matrix for predicting unseen
preference\cite{su:survey,adomavicius:survey}. It can be categorized
into memory-based CF, which contains the whole matrix on memory, and
model-based CF, building a model for
estimation\cite{breese:empirical}. The most effective memory-based
algorithms known so far is item-based CF\cite{sarwar:item}.
Recently, making use of matrix factorization, a kind of model-based
approach\cite{paterek:svd, rennie:fmmm, salakhutdinov:pmf, lee:nmf,
yu:npca}, is known as the most efficient and accurate, especially
after those approaches won the Netflix prize in 2009. Content-based
methods, on the other hand, recommend items based on their
characteristics as well as specific preferences of a
user\cite{jannach:rs}. Pazzani\cite{pazzani:content} studied this
approach in depth, including how to build user and item profiles.
Last category, hybrid approach, tries to combine both collaborative
and content-based recommendation. Koren\cite{koren:multifaceted}
suggested effectively combining rating information and user, item
profiles for more accurate recommendation.

Recommender systems have concentrated on recommending media items
such as movies, but recently they are extending to academy. Most
popular application is citation recommendation\cite{he:contextcite,
mcnee:cite, tang:topiccite, strohman:cite}. Recently,
Matsatsinis~\cite{matsatsinis:paper} introduced scientific paper
recommendation using decision theory. Sugiyama\cite{sugiyama:paper}
extended scholarly paper recommendation with citation and reference
information.

Although recommender systems are very popular in commercial
applications these days, it is still difficult to evaluate them due
to the lack of standard methods. Traditional recommender
systems~\cite{goldberg:tapestry, resnick:grouplens, hill:vc} were
usually introduced in Human-Computer Interaction community, so they
have been evaluated by user study. This approach is still used,
especially for verifying improvement in terms of user experience.

\section{Methodology}

\begin{figure}[h]
    \centering{}
    \includegraphics{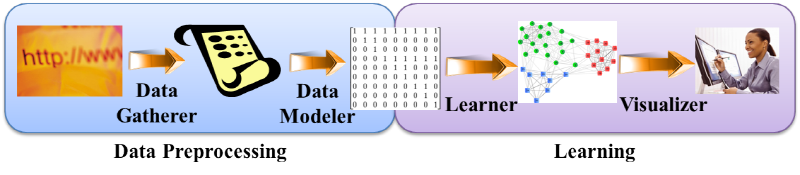}
    \caption{Recommendation Flow}
    \label{fig:arch}
\end{figure}

Figure \ref{fig:arch} shows the flow of our system. First, our
system gathers data and preprocess it, by applying Bag-of-word model
to the corpus. In actual learning process, we apply lazy learning
method similar to k-Nearest Neighbors (kNN). Thus, we estimate
preference of a target user and recommend the most preferred papers,
when the system gets a query, specifying the target user. For this
task, we applied clustering and neighbor-based recommendation
algorithm. Finally, the result is conveyed to the user by
visualizer. In this section, we describe each component in detail.

\subsection{Data Gatherer}
To get information of research papers, we have implemented a web
data gatherer for two research paper search engines and storage:
IEEE Xplore and ACM Digital Library.

IEEE Xplore Digital Library provides a unique web page for
conferences or journals, of a specific year or edition respectively.
On each page, there is a list of published papers, so we retrieve
the list of URLs using regular expression matching. After getting
the URL list, our data gatherer retrieves the information of all
research papers in the list by iterating each URL in the list.

ACM Digital Library also provides a unique web page for conferences
or journals published by ACM. The data gatherer finds a list of URLs
which have information for each research paper using regular
expression matching. Unlike IEEE Xplore, since the single page
includes URLs of all the papers, our data gatherer does not need to
traverse multiple pages for one conference or journal. With the URL
list, the data gatherer gathers the information of all research
papers in the list by iteratively visiting each URL in the list.

We have experienced a challenge while developing the data gatherer.
There are different representations for same researcher's name. To
solve this problem, we have developed several rules to handle any
orders of first name, last name, middle name and suffix. Also, we
have implemented a function to infer authors' full name using the
co-occurrence with other authors.

\subsection{Data Model}
\subsubsection{Bag-of-word model}
With the gathered data, we modeled them by a bag-of-word model. In
this model, each word appeared in the whole document corpora becomes
an attribute. Then, each document is represented by a bit vector,
indicating whether each word appears or not. This model is based on
two assumptions; 1) word probabilities for one text position are
independent of the words that occur in other positions (Naive Bayes
Assumption) and 2) the probability of encountering a specific word
is independent of its position. (Independent Identical Distribution
Assumption) This assumption is incorrect, but it is known that this
does not seriously affect classification or learning
task.~\cite{mitchell:ml} We combined title, key words, and abstract
to construct a set of words representing a paper.

\subsubsection{Heuristics}
For more efficient processing, we applied some heuristics. First, we
removed stop words such as "the" or "of". These words appear in
almost every document in English, so they are not useful for
classifying or filtering some specific documents, but just slow down
computation speed by increasing the text length. We removed about
140 words which were selected manually. This process reduces the
length of dictionary, resulting in reduced dimension of the
clustering work, so we expect speed improvement.

The other heuristic applied is stemming. In English, same word can
be used as different parts, usually in a slightly different form.
For example, "clear", "clearly", and "cleared" have same meaning,
but used in different forms for its position or role in the
sentence. It is much better to deal with these minor changes of
forms as same words, as it can dramatically reduce the dimension.
However, this work is not straightforward. As a first step, we just
removed last "ed", "ly", and "ing" from the word, whenever
encountered.

\subsection{Learner (Recommender)}
Using the crawled documents and data model discussed so far, we are
ready to proceed to our main goal: personalized recommendation of
academic papers. As a perspective of recommendation system, we can
consider authors as users and papers as items. We will use these
terms interchangeably henceforth. We can think of recommendation
system as a task to fill out missing preference data on a user-item
matrix, based on observed values. There can be lots of schemes to
decide proper values for missing preference. Filling with the user's
average or item's average can be a simple baseline. In this section,
we discuss fundamental characteristics of our problem, and then
describe our algorithm.

\subsubsection{Inherent Characteristics of Problem}
The information we gather contains each paper's title, list of
authors, key words, and abstract. In order to build a user-item
matrix with this data, we assume that users are interested in their
own papers. Thus, we set high score (in this paper, 5) to every
$<$author, paper$>$ pair that the paper is written by the author. We
use 1-5 scale as it is widely used in recommendation systems in
literature.

We claim that this user-item matrix we use is extremely sparse,
which means most of values are missing while only small portion of
them are observed. This situation is common in recommendation,
though. According to Netflix Prize data, only 1\% of cells of the
user-item matrix are observed values. Nonetheless, it has been shown
that it is possible to accurately estimate missing data only using
small amount of observed data. In our situation, however, the
sparsity can be worse. Regularly, one author writes only one or two
papers in one conference proceeding. There are only at most two or
three top-level conferences in each field, the maximum number of
papers one author can publish a year is about 10. This is an ideal
case, and most researchers may have only one or two papers. Thus,
our matrix have only a few number of preference data.

More serious problem is that we do not have "dislike" information.
When we request users to explicitly rate items in a common
recommendation system, we can get both positive and negative
feedback from the user. For example, we can get "very like" feedback
for the movie "Titanic" as well as "very hate" one for the "Shrek
2." Based on this variety, we can infer that the user may prefer
romantic movies to animations. In our data, however, we do not have
negative feedback. This problem makes difficult us to use
widely-used collaborative filtering algorithms.

\subsubsection{Naive Recommender}
We basically assume that authors will like papers similar to ones
they wrote before. In this context, we note that similar papers mean
ones dealing with similar topic. In our Naive Recommender, we just
apply this assumption. When we try to recommend a set of papers to a
specific user, we first calculate similarity between every paper and
the user's own papers. Then, we take the highest similarity as the
score of that paper. This process is similar to k-Nearest Neighbors
(kNN) algorithm. That is, we can easily select and recommend most
similar $n$ papers to the target user's previous paper. We used
vector cosine of our data model (bit vector) as the similarity
measure.

\begin{figure}[h]
\centering{}
\includegraphics{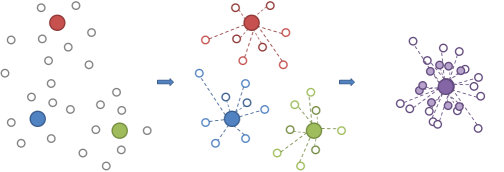}
\caption{Visualization of Clustering and kNN} \label{fig:cluster}
\end{figure}

However, the real situation is a little bit more complicated, as the
user may have written more than one paper. It is still kNN, but we
can have more than one queried point. Thus, we applied clustering
first. All candidate papers are assigned to only one of the most
similar paper written by the target user. This process is similar to
K-means, but the centroids are also papers, so their geometric
location in the space cannot change. Thus, we do not need to iterate
in our case. After assigned to a cluster, the score is calculated
based on the distance between the candidate paper and its centroid.
For example, as shown in Figure \ref{fig:cluster}, each big circle
represents a centroid of a cluster and small circles connected to
the centroid are members of its cluster. Using the calculated score
as a distance metric for kNN, we select $k$ papers for
recommendation to the target user.

To illustrate, assume that the user $u$ wrote only two papers ($x_1$
and $x_2$, respectively) until now. We calculate estimated
preference of new paper $p$ by the user $u$. First, we calculate
similarity between $p$ and $x_1$. Let's say this similarity as
$s_1$. We also calculate similarity between $p$ and $x_2$, namely
$s_2$. Then, we compare $s_1$ and $s_2$. We set estimated preference
of $p$ by user $u$ as $max(x_1, x_2)$. This is because the user may
like a paper when it is related to his one of the interested topics.
Although it is not related to other papers, the author may still
like it if it is related to at least one of the topic in which he is
interested. More formal formula for estimating preference of user
$u$ for item $i$ is given as:
\begin{equation}
max_i {( { {sim(x, i) \times 5} \over {max_{i,k} sim(j, k)} } )}
\end{equation}
where, $x$ is index of papers which the user $u$ published, and $i,
j, k$ are index of all papers. The score is scaled to 5 as we set 5
points for authorship. This scale will be lowered for referenced
papers.

\section{Evaluation}

\subsection{Classification Accuracy}
First, we measured accuracy of our system by the following simple
classification scheme. As we crawled papers (and authors) from three
different areas (ML, HCI, DB), we observed how many papers are
actually recommended from the researcher's own area, when the corpus
is mixed. For this experiment, we recommended 10 papers each among
the overall 10,386 papers to 10 ML researchers, 10 HCI researchers,
and 10 DB researchers. This task can be seen as a kind of
classification of each researcher, based on the papers recommended
to them. The result is shown in Figure \ref{tab:classification}. In
overall, our system recommended papers from correct area with
accuracy of 89\%. One thing to note is that, however, recommendation
from different area may not an incorrect result, as some researchers
actually do research cross over other areas. For example, robotics
research is highly related to ML as well as HCI. Thus, fine detail
of evaluation should be conducted with real users, as described in
the next subsection.

\begin{table}[!h]
\centering \caption{Classification Accuracy}
\label{tab:classification}

\begin{tabular}{c|c|c|c|c}
\hline Area & ML Paper & HCI Paper & DB Paper & Accuracy
\tabularnewline \hline \hline ML Researchers & 84 & 0 & 16 & 84\%
\tabularnewline \hline HCI Researchers & 3 & 88 & 9 & 88\%
\tabularnewline \hline DB Researchers & 4 & 1 & 95 & 95\%
\tabularnewline \hline
\end{tabular}
\end{table}

\subsection{User Study Design}
Our system is aimed to recommend similar papers to the target user's
previously published papers, assuming that researchers will like
similar papers to their previous research topics. In order to verify
whether this assumption is true, we evaluated the content relevance
between the target user¡¯s previous published papers and the papers
recommended by our system. We conducted focus group user study by
interviewing three professors from all different fields of Computer
Science area, i.e., Machine learning, Data Base, and Human-Computer
Interaction. One professor is a junior professor, and the other two
are senior professors. The data we used in this evaluation is
summarized in Table \ref{tab:papersummary}.

\begin{table}[!h]
\centering
\caption{Crawled Data}     
\label{tab:papersummary}

\begin{tabular}{c|c|c|l|c}
\hline Area & Paper & Author & Conferences (Years) & Time (sec)
\tabularnewline \hline \hline
ML  & 3,644 & 5,786 & ICML(04-09), KDD(04-10), COLING(04-10), & 38.8   \\
    &       &       &   UAI(04-09), SIGIR(04-10), JML(04-11) &
\tabularnewline
\hline
HCI & 2,557 & 4,728 & CHI(03-09), ASSETS(02-09), CSCW(04-11), & 33.5    \\
    &       &       & Ubicomp(07-10), UIST(04-10) &
\tabularnewline
\hline
DB & 4,156 & 7,213 & ICDE(06-10), SIGMOD(06-10), VLDB(06-10),   & 92.2    \\
        &       &       & EDBT(08-11), PODS(06-10), CIKM(06-10) &
\tabularnewline
\hline
\end{tabular}
\end{table}

We provided 10 papers, to each participant, recommended by our
system. The survey listed title, authors, proceedings name, and
abstract of those papers. The participants are asked to read and
indicate how the papers are relevant to their research. We used a
Likert-scale between 1 (not relevant at all) and 6 (perfectly
relevant), in order to prevent voting to middle-way. After
evaluating all 10 papers, we asked how much the list of papers was
relevant to the researcher¡¯s previous research and current research
separately. One more thing we wanted to evaluate is the usefulness
of recommended papers. We asked the subjects to indicate the number
of papers they would take time to read and get useful information
among 10 recommended papers. Lastly, we asked how they are satisfied
with the system in overall and how much they are willing to use the
system. Here, we also used 1-6 Likert-scale question.

\subsection{Result}

As shown in Table~\ref{tab:userstudy}, three subjects indicated
recommended papers tend to relate to their previous and current
research. When we asked how the recommended paper list is related to
their previous research in overall, all gave higher than 5 point.
(5.5, 5.0, 5.5) However, for the relevance of their current research
topic, even though two professors gave 5.0 and 5.5, respectively,
the other gave 2 point. In this case, she has worked on so many
various topics before, so our system recommended papers that are
relevant to every topic she has been interested in. In this way, for
her, among 10 recommended papers there were only two papers related
to current research topics. Overall, all the professors were
satisfied with the results of the recommended papers in respect of
the topic relevance to their research.

For senior professors, our system recommended four papers that their
previous students published. Since their previous students now
graduated, they did not publish the recommended papers together with
the professor, but the paper¡¯s topics were relevant to what they
have done with the professor. The fact that our system is able to
recommend these papers also ensures good performance of our system
in recommending relevant topic¡¯s papers.

When we asked the subjects the number of papers to read, they
replied, realistically they would read only the papers that are
highly related to their research, which was about two papers among
10 papers since they do not have enough time to read all the papers.
This result is also very satisfied, when we consider the fact that
we did not use any other user¡¯s profile information and just only
used content based on their previous research. Lastly, all of the
subjects marked 6.0 point out of 6.0 to use this recommendation
system, indicating that our research is valuable for real users.

\begin{table}[!h]
\centering \caption{Relevance of recommended papers}
\label{tab:userstudy}
\begin{tabular}{l|c|c}
\hline Subjects & Average & Standard Deviation \tabularnewline
\hline \hline Subject 1 & 4.40 & 1.26 \tabularnewline \hline Subject
2 & 4.00 & 1.56 \tabularnewline \hline Subject 3 & 3.25 & 1.72
\tabularnewline \hline Total & 3.88 & 1.52 \tabularnewline \hline
\end{tabular}
\end{table}

\section{Discussion}

\subsection{Contribution and User Interface}
\begin{figure}[h]
\centering{}
\includegraphics{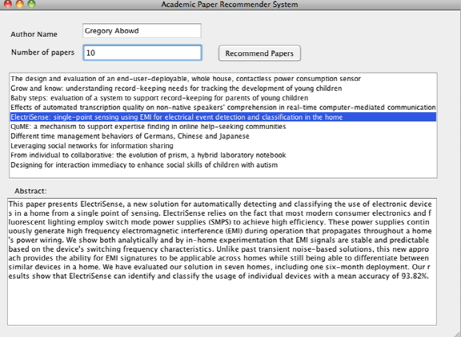}
\caption{Graphical User Interface Prototype} \label{fig:ui}
\end{figure}

In this study, we investigated the way to reduce academic
researchers' labor-intensive workload, going through all the
conferences and journals to find out any scholarly papers that might
relate to their research topic. Our system provides simple Graphical
User Interface (GUI), shown in Figure~\ref{fig:ui}, that requires
two input fields, name of researcher and the number of academic
papers the user wants to be recommended.

\subsection{Limitation and Future Work}
Even though our system showed good performance on recommending
relevant topic's paper, we identified two limitations on our
content-based recommendation. First, we cannot distinguish the
meaning of topics that are narrowed by few specific words. In other
words, our system recommends papers based on the words' frequency,
so our system will recommend papers that contain many words that the
user may be interested in. This cannot discover few words that
restrict meaning of its topic, so it causes recommending the paper
that has not relevant topic to the user. Also, there are users who
have worked previous study on various topics, but are not interested
in anymore. Also in this case, because our system does not have any
additional information about whether the user is still interested in
the paper or not, it is hard to distinguish the papers that
recommend to users. We may be able to extend by applying publication
year in some way.

To overcome these limitations we need to recommend papers based on
not only relevant topics but also other user information. We suggest
obtaining a user input about whether they like the recommended
papers or not would be helpful information to differentiate the
papers that users would be interested in more accurately. Also,
through the focus group interview we discovered the interesting fact
that even though the topics are not as much as relevant to their
research topic, they showed great interest to the papers that their
peer researchers, i.e., their former students or the researchers
they have done research together before, wrote. In this way, it will
be important to include the information about relevant researchers
to users and recommend papers that they found interesting or they
have wrote. Also, the subjects replied, if we provide information
about which researcher liked this papers, it would also give them
great reason and motivation to read that paper.

For the perspective of machine learning, we may need to consider
about scalability. Although our current system runs within a few
minutes, it may take more time when we crawl more data. First, we
can improve accuracy of similarity measure by allowing counting the
frequency of each word in a document, instead of bit vector model.
TF-IDF model~\cite{manning:ir} can be a great candidate to
implement. In this model, we give more weight for frequently used
words in a specific document, but not in other ones. Also, we may
need to speed up the calculation. For this, dimension reduction will
be helpful. Specifically, it would be better to add more stemming
logic because this can deal with more words as same ones, so we can
successfully reduce dimension. We may use L-Distance
algorithm~\cite{ldistance} for calculating similarity of each word
pair, and decide whether they are same or not.

\section{Conclusion}
In this paper, we have presented a Personalized Academic Research
Paper Recommendation System, which recommends related articles for
each researcher. Thanks to our system, researchers can get their
related papers without searching keywords on Google or browsing top
conferences' proceedings. Our system makes three contributions.
First, we have developed a web crawler to retrieve a huge number of
research papers from the web. Second, we define a similarity measure
for research papers. Third, we have developed our recommender system
using collaboration filtering methods. Evaluation results show the
usefulness of our system.

\small{
\bibliographystyle{abbrv}
\bibliography{ref}
}


\end{document}